\documentclass[useAMS,usenatbib]{mn2e}
\usepackage{graphicx}
\usepackage{times}
\usepackage{amssymb}
\usepackage{natbib}
\usepackage{epsf}
\usepackage[dvips]{epsfig}
\def\gsim { \lower .75ex \hbox{$\sim$} \llap{\raise .27ex \hbox{$>$}} }
\def\lsim { \lower .75ex \hbox{$\sim$} \llap{\raise .27ex \hbox{$<$}} }

\newcommand{\apj}{ApJ}
\newcommand{\apjl}{ApJL}
\newcommand{\apjs}{ApJS}
\newcommand{\aj}{AJ}
\newcommand{\mnras}{MNRAS}
\newcommand{\nat}{Nature}

\newcommand{\araa}{ARA\&A}
\newcommand{\aapr}{A\&ARv}
\newcommand{\aap}{A\&A}

\begin{document}

\title[On the genealogy of the Orphan Stream] 
{On the genealogy of the Orphan Stream}

\author[Sales et al.]{
\parbox[t]{\textwidth}{
Laura V. Sales$^{1}$, 
Amina Helmi$^{1}$, 
Else Starkenburg$^{1}$, 
Heather L. Morrison$^{2}$,
Ethan Engle$^{2}$,
Paul Harding$^{2}$,
Mario Mateo$^{3}$, 
Edward W. Olszewski$^{4}$,
and Thirupathi Sivarani$^{5}$} 
\vspace*{6pt} \\
\\
$^{1}$ Kapteyn Astronomical Institute, University of Groningen, P.O Box 800, 9700AV, Groningen, The Netherlands. \\
$^{2}$ Department of Astronomy,Case Western Reserve University, Cleveland OH 44106-7215, United States.\\
$^{3}$ Department of Astronomy, University of Michigan, Ann Arbor, MI 48109, United States.\\
$^{4}$ Steward Observatory, University of Arizona, Tucson, AZ 85721, United States.\\
$^{5}$ Department of Physics \& Astronomy and JINA: Joint Institute for Nuclear Astrophysics, Michigan State University
}

\maketitle

\begin{abstract}
We use N-body simulations to explore the origin and a plausible orbit
for the Orphan Stream, one of the faintest substructures discovered so
far in the outer halo of our Galaxy.  We are able to reproduce its
position, velocity and distance measurements by appealing to a {\it
single} wrap of a double-component satellite galaxy. We find that the
progenitor of the Orphan Stream could have been an object similar to
today's Milky Way dwarfs, such as Carina, Draco, Leo~II or Sculptor;
and unlikely to be connected to Complex A or Ursa Major II. Our models
suggest that such progenitors, if accreted on orbits with apocenters
smaller than $\sim 35$ kpc, are likely to give rise to very low
surface brightness streams, which may be hiding in the outer halo and
remain largely undetected with current techniques.  The systematic
discovery of these ghostly substructures may well require wide field
spectroscopic surveys of the Milky Way's outer stellar halo.
\end{abstract}


\section{Introduction}
\label{sec:intro}

Tidal streams represent direct signatures of the merging history of
galaxies. In Cold Dark Matter (CDM) models, structures grow in a
bottom-up fashion, by the accretion of smaller sub-units. After
entering in the gravitational domain of a larger system, tides
effectively remove material from the satellites, creating ``tails'' of
particles, either dark matter, gas or stars that approximately trace
the orbits of their progenitors. The existence of tidal streams in the
halos of galaxies would therefore be a natural expectation in the
hierarchical paradigm of structure formation.

However, it is still unclear how fundamental or dominant
mergers have been in the build-up of our Galaxy. In a recent paper,
\citet{bell08} found that the amount of substructure in 
the Milky Way's halo is consistent with the most extreme
scenario, in which it was {\it entirely} formed from the
accretion of satellites.

Although a large amount of stellar streams are predicted in these
hierarchical models
(\citealt{johnston95,johnston98,helmi99,bullock01b,helmi03b,bullockandjohnston05}),
detection is generally difficult due to their low surface brightness
and low contrast against Galactic field stars.  This implies that our
own Galaxy and also our closest neighbour, M31, may offer the best
chances for identifying streams in stellar surveys. Perhaps the
clearest and best known example is the Sagittarius Stream, discovered
about a decade ago.  With the advent of large wide-field surveys like
2MASS \citep{2mass} and the Sloan Sky Digital Survey
\citep{york00,sloandr5} it was possible to trace this stream out to
360$^\circ$ on the sky \citep{majewski04}, allowing the modelling of
the orbit and intrinsic properties of its progenitor, the Sagittarius
dwarf \citep{helmi01,
ibata01b,martinez_delgado04,helmi04,fellhauer06}.  Also M31 shows a
prominent ``Giant Arc'', a stellar stream of average surface
brightness $\Sigma_V \sim 30 \rm mag/arcsec^2$, whose progenitor has
not yet been confirmed
\citep{ibata01,guhathakurta06,kalirai06,fardal06b,font06c, gilbert07}.
Therefore, these surveys have consolidated the idea that the stellar
halos of M31 and the Galaxy may well be highly lumpy components. This
appears to also be the case for galaxies beyond the Local Group, as
recent studies have shown
\citep[e.g.][and references therein]{martinez_delgado08b}.

\begin{center}
\begin{figure*}
\includegraphics[width=0.32\linewidth,clip]{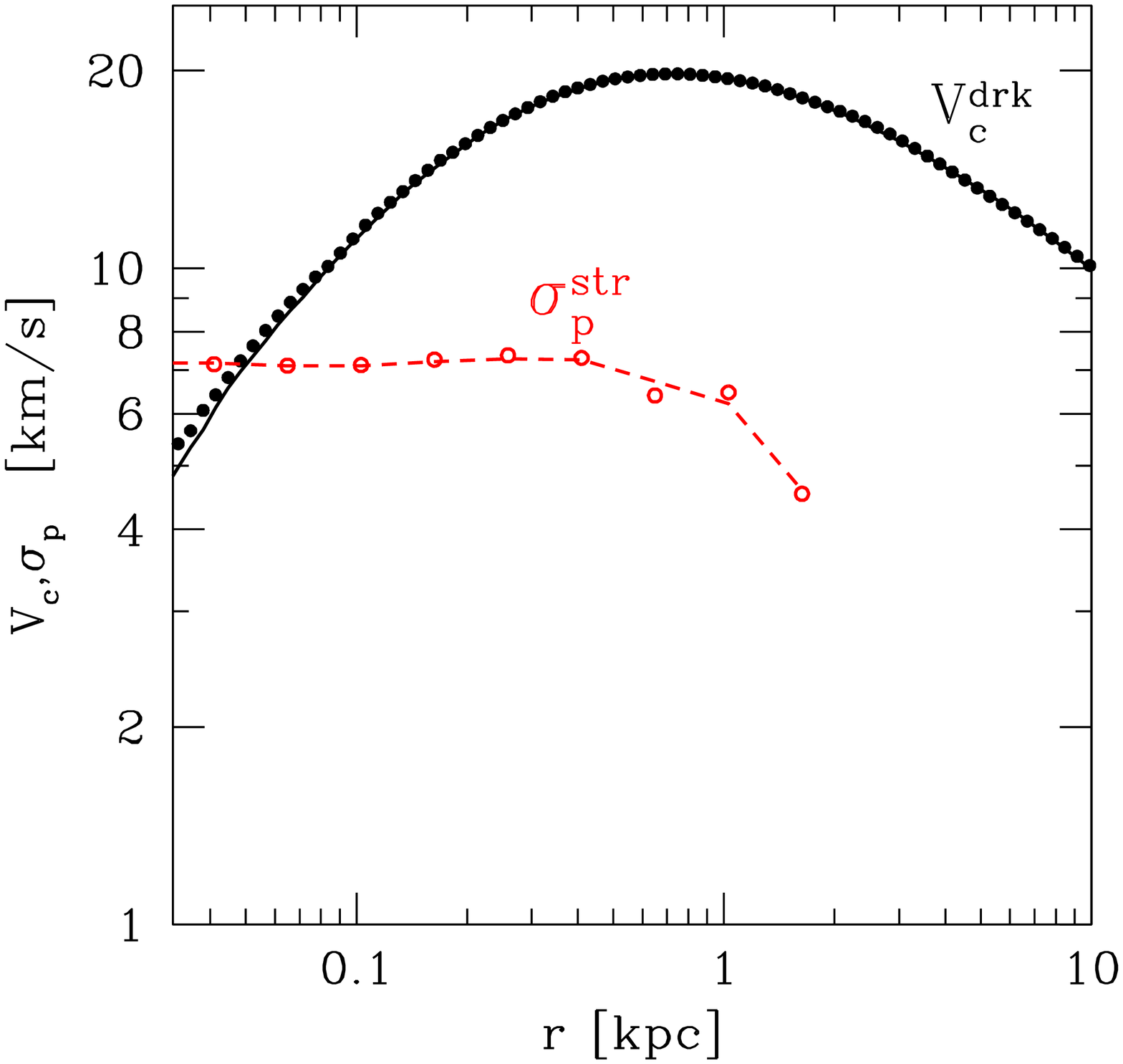}
\hspace{0.2cm}
\includegraphics[width=0.32\linewidth,clip]{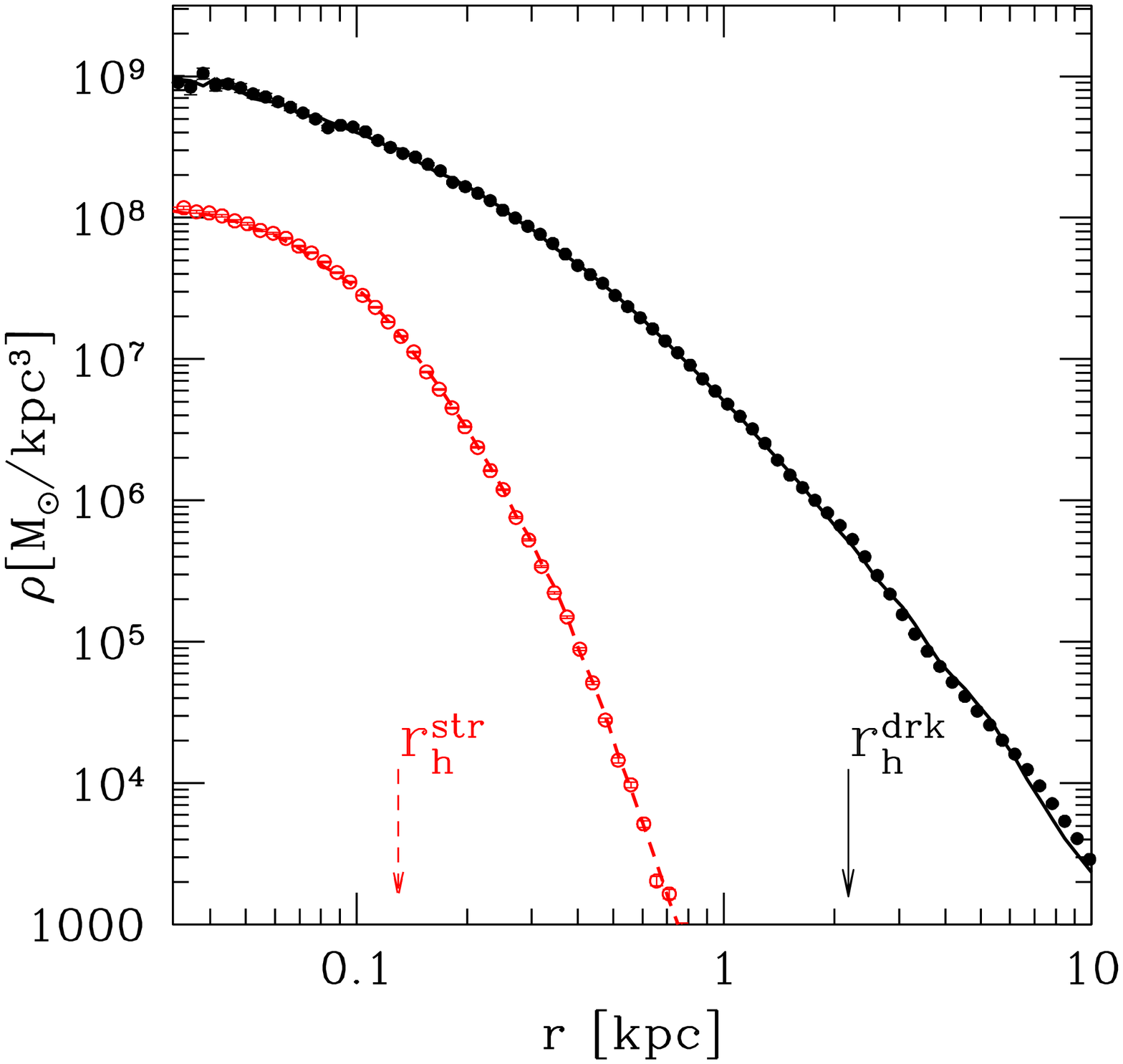}
\hspace{0.2cm}
\includegraphics[width=0.32\linewidth,clip]{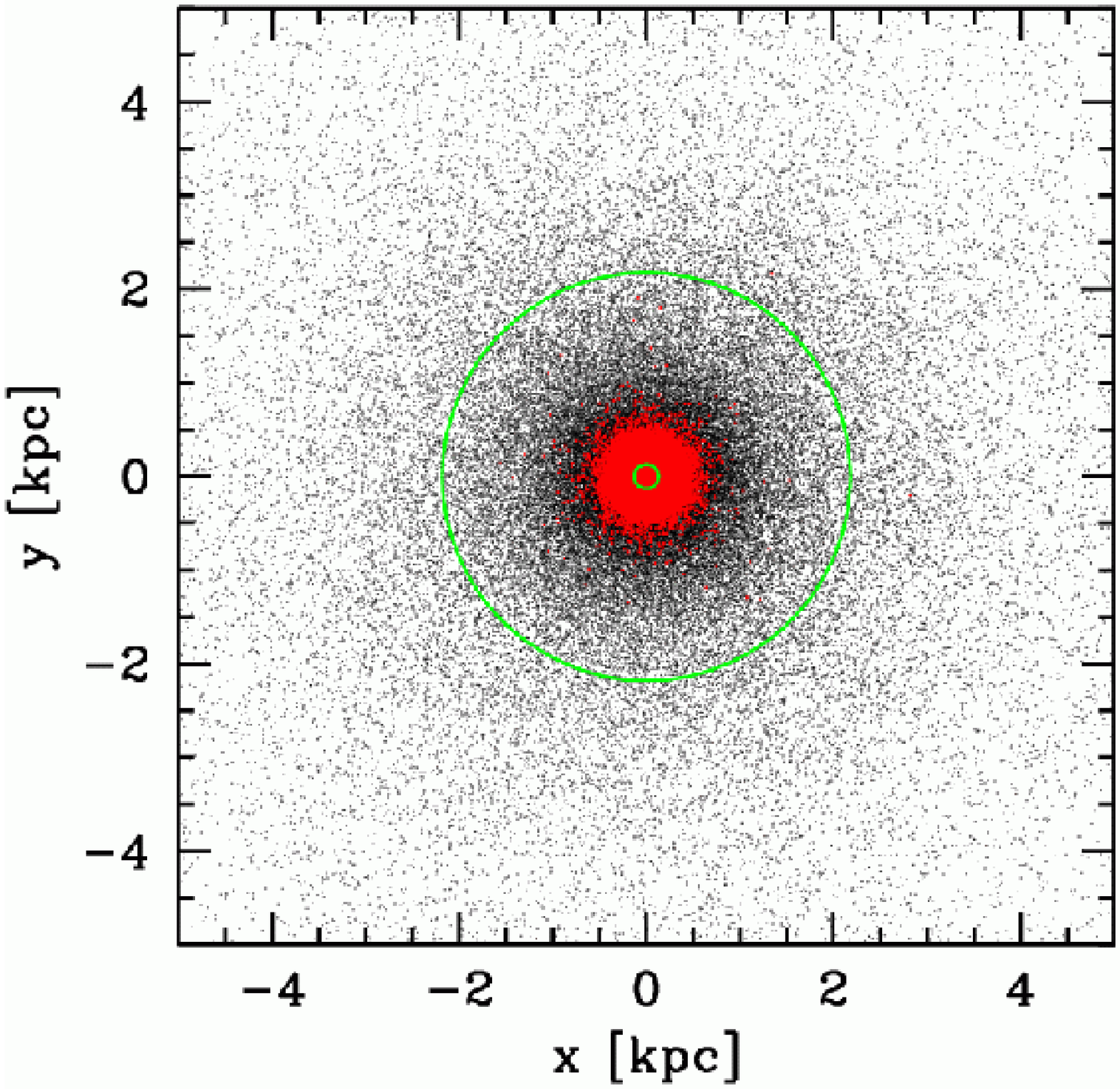}
\caption{ Properties of the satellite model considered in this work.
{\it left:} dark matter circular velocity (black squared dots, solid
line) and the (projected) velocity dispersion of the luminous
component (open red circles, dashed line) of the satellite as a
function of radius. Lines indicate the profiles as set up in the
initial conditions, while points show these quantities after the model
is relaxed in isolation during 2 Gyr.  {\it middle:} dark and luminous
matter density profiles for the satellite, arrows show the half-mass
radii for each component.  Lines and points are colour-codded as in
the previous panel.  {\it right:} projected positions of the dark
matter (black) and luminous (red) particles in the relaxed satellite
model. The green circles indicates the half-mass radius, highlighting
the appreciable segregation of the stars with respect to the dark
halo.  }
\label{fig:relax}
\end{figure*}
\end{center}

Stellar streams are useful tracers of their progenitor's orbit, due to
their coherence in phase-space. This property allows us in some cases to
link a given tidal feature with its parent satellite; for instance, by
extrapolation of its great circle on the sky \citep{lyndenbell95}. At
the same time, tidal streams also hold essential clues about the
object in which they originated, as their luminosity and cross section
are directly related to the mass and velocity dispersion of their
progenitors \citep{johnston98}. Stream properties also depend on the
time of disruption, since the material that is stripped off earlier in
time tends to become broader, and hence, to have lower surface
brightness as time goes by \citep{helmi99}.  In this context, we
naturally expect an observational bias towards detecting remnants of
{\it recent} accretion events involving {\it massive} progenitors, as
the case of the Sagittarius stream in the Milky Way or the Giant Arc
in M31.  However, an unusually narrow and faint stream discovered in
our Galaxy stands out from this general trend. The Orphan stream
\citep{grillmair06,belokurov07b} with only $\sim 0.70$ kpc of full
width half-mass (FWHM) projected on the sky, is about $\sim 5$ times
narrower than the Sagittarius Stream, and has a surface brightness
which is about a factor $\sim$2 lower. Given its atypical properties,
the Orphan Stream provides us with the opportunity of studying the
fainter end of the family of objects that built up the stellar halo of
our own Galaxy.

As highlighted by its name, the Orphan Stream progenitor has not yet
been identified, although its properties (mainly cross-section and
luminosity) seem to favour a dwarf-like object rather than a globular
cluster.  Some attempts have been made to link this stream to other
known objects of our Galaxy. Belokurov et al. (2007b), based on the
paths defined by great circles on the sky (a technique first developed
by \citealt{lyndenbell95}), highlighted a potential connection of the
stream with several globular clusters (Palomar 1, Arp2, Terzan 7 and
Segue 1), as well as to the recently discovered faint dwarf galaxy,
Ursa Major II (UMa~II, \citealt{zucker06b}). In a later paper,
Fellhauer et al. (2007) used numerical simulations to test the likely
association between the Orphan Stream and UMa~II. These authors propose
a model in which both objects have a common origin, also allowing for
a physical association of the stream with a set of high velocity
clouds named Complex A. However, this model requires the stream to be
the result of the exact overlap on the sky of two independent wraps,
which at face value appears somewhat contrived given its noticeable
cohesion.

In a novel approach, \citet{shoko07} exploited the possible relation
of Complex A with the stream to estimate its average
distance. However, the method under-predicts the distances by a factor
of $\sim 3$, also disfavouring the scenario in which the Orphan Stream
and Complex A share the same orbit.

This paper presents a new attempt to find a plausible orbit for the
Orphan stream progenitor, that is able to reproduce the current
position and velocity measurements appealing to only one wrap; as
suggested by the coherence observed in the images of the stream. Our
model for the progenitor aims to be consistent with dwarf
spheroidal galaxies, like those found orbiting the Milky
Way today. The paper is organized as follows: we describe the simulations as
well as the progenitor model in \S~\ref{sec:numexp}, we present the
results and discussion in \S~\ref{sec:analysis} and
\S~\ref{sec:predictions}.  Finally, we summarize our main conclusions
in \S~\ref{sec:conc}.

\section{Numerical Modelling}
\label{sec:numexp}

We use the N-body code GADGET-2 \citep{gadget2}
{\footnote{http://www.mpa-garching.mpg.de/$\sim$volker/gadget/index.html}} 
to study the evolution of a two-component satellite galaxy
orbiting in the (fixed) gravitational potential of the Milky Way. 
We model the Galactic potential as follows:

\begin{itemize}
\item a spherical NFW \citep{nfw96,nfw97} dark matter halo:

$$\rho(x=r/r_{vir}) \propto \frac{1}{x(1+cx)^2}$$

with mass $M_{vir}=1\times10^{12} \rm M_\odot$, concentration $c=12$ and
virial radius $r_{vir} = 258$ kpc \citep{klypin02}, 

\item a Hernquist bulge:

$$\rho(r) = \frac{M_{blg}a_b}{2\pi r}\frac{1}{(r+a_b)^3} $$

with mass $M_{blg}=3.4\times10^{10} M_\odot$ 
and scale length $a_b=0.7$~kpc, 

\item a Miyamoto$-$Nagai disk:

$$\rho(R,z)=\frac{b^2M_{dsk}}{2\pi} \frac{aR^2+(a+3\sqrt{z^2+b^2})(a+\sqrt{z^2+b^2})^2}{[R^2+(a+\sqrt{z^2+b^2})^2]^{5/2}(z^2+b^2)^{3/2}}$$

with parameters: $M_{dsk}=1\times10^{11} M_\odot$, $a=6.5$ kpc and $b=0.26$ kpc
(Johnston et al. 1999).\nocite{johnston99}
\end{itemize}

The composite circular velocity at the solar distance is $V_c = 220$
km/s in agreement with observations. The circular velocity at the
virial radius is $V_c(r_{vir})=V_{vir}=136$ km/s.  Note that for
simplicity we have assumed a spherical dark matter halo, and that the
mass distribution is time-independent. This assumption is justified
because the gravitational potential inside the orbit of the Orphan
stream ($< 40$ kpc, see next Section) is unlikely to have changed
significantly in the recent past because of the presence of a
relatively old thin disk. On the other hand \citet{penarrubia06b}
suggest that the properties of tidal streams mainly reflect the {\it
present-day} potential of the primary halo and are not fundamentally
affected by its growth in time.

Our model of the satellite has two spherical components: an extended
dark matter halo and a more concentrated ``luminous'' component (see
middle and right panels of Figure \ref{fig:relax}).  Dark matter
particles are distributed following a Hernquist profile of mass
$M_{drk}^{sat}=2.5\times10^8 M_\odot$ and characteristic scale
$a_{sat}=0.9$ kpc. The circular velocity of the dark halo peaks at
$r_{max}\sim 0.75$ kpc reaching $V_{max}\sim 20$ km/s (see left panel
of Figure \ref{fig:relax}), and hence would be consistent with the
properties of the Milky Way dwarf spheroidals (Pe\~narrubia et
al. 2007,2008)\nocite{penarrubia07,penarrubia07b}.  The second component
mimics the stellar content of a dwarf galaxy and is represented by a
Plummer profile of mass $M_{str}^{sat}=7.5 \times 10^5
M_\odot$ and characteristic scale $b_{sat}=0.1$ kpc. The half light
radius of our model satellite is $r_h \sim 0.13$ kpc and its central
velocity dispersion is $\sigma_0 \sim 7$ km/s.

We used the web-tool BaSTI
{\footnote{www.te.astro.it/BASTI/index.php}} to generate isochrones
and convert the stellar mass ($M_{str}=7.5 \times 10^5 M_\odot$) to
the luminosity of our modelled satellite. Given the typically old
stars present in dwarfs galaxies, and assuming a single population of
10 Gyr and [Fe/H]$\sim -2$ dex, the conversion factor we obtain is
$\gamma \sim 2.9$. This gives a total luminosity for the progenitor in
our simulations of $L \sim 2\times10^5 L_\odot$, consistent with the
luminosity of dwarf spheroidals in the Local Group. This object is
dark matter dominated, with a mass-to-light ratio $\gamma \sim 39$
measured within the half light radius. With these parameters the
model is comparable to several classical dwarfs, such as Carina,
Leo~II or Sculptor.

The number of particles used in our simulations are 5$\times10^5$ and
2$\times10^5$ for the dark and stellar components, respectively.  We
use an unequal softening scheme, with a Plummer-equivalent softening
length: $\epsilon_{DM}=0.04$ kpc and $\epsilon_{str}=0.004$ kpc for
the dark matter and luminous component, respectively.  We set up the
initial conditions following the procedure outlined by
\citet{hernquist93}, aimed to generate multi-component systems in
dynamical equilibrium. The satellite is first evolved in isolation
during 2 Gyr ($\sim 200$ crossing times for a particle at the half-light
radius, while for one at the edge, i.e. at 5 kpc from the center, this
corresponds to $\sim 5$ crossing times),
where it is allowed to relax. Then, the satellite was put on its
current orbit in the fixed Milky Way potential.

As stated above, our model for the Orphan stream progenitor could well
represent one of the ``classical'' Milky Way dwarf galaxies, where
``classical'' is used to distinguish the first eleven discovered
satellites around our Galaxy \citep{mateo98,vandenbergh99}, from the
more recently identified SDSS dwarfs
(\citealt{zucker04,willman05b,martin06,belokurov06}, Zucker et
al. 2006a,b, \citealt{belokurov07,irwin07,
majewski07,simon07})\nocite{zucker06,zucker06b}.  These new dwarfs are
typically $\sim 100$ times less luminous, but with comparable half
light radii ($\sim 80-500$ pc) and velocity dispersion ($\sim 4-8$
km/s). Therefore their surface brightness is appreciably lower ($\sim
28-30$ mag/arcsec$^2$) and their mass-to-light ratio is $\sim 10-50$
times higher. One of the strongest constraints on the Orphan Stream
progenitor comes from its total luminosity. Belokurov et al.  (2007b)
estimate that the total $r$-band absolute magnitude for the (visible)
portion of the stream is $M_r\sim-6.7$. Since leading and trailing
arms should have comparable mass, a lower limit to the total luminosity is
roughly $L \sim 10^5 L_\odot$, ruling out practically all new SDSS
dwarfs as suitable candidates \footnote{Only the Canes Venatici I dwarf,
whose luminosity $L\sim1.2 \times 10^5 L_\odot$ is above this lower
limit. However, this galaxy is too extended ($r_h=0.56$ kpc) to give
rise to a stream as narrow as observed}.

\begin{center}
\begin{figure}
\includegraphics[width=84mm]{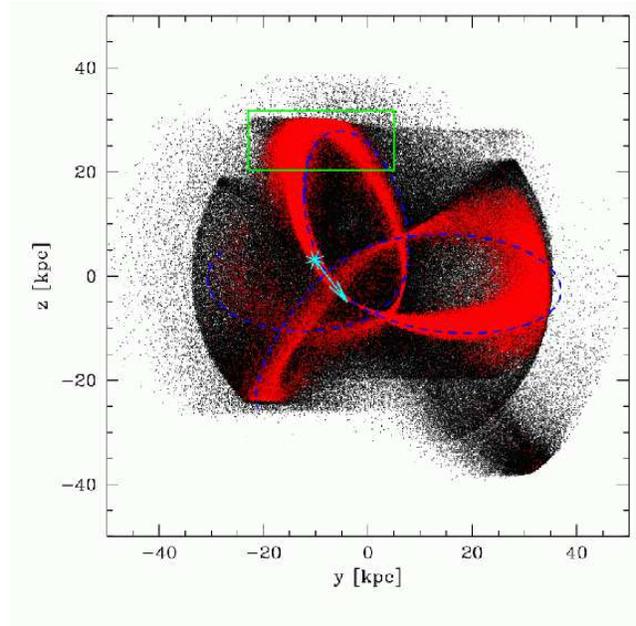}
\caption{Distribution of dark matter (black) and star particles (red)
at $5.3$ Gyr after infall. The section matching the Orphan Stream
measurements is enclosed by the green rectangle. It sits on the {\it
trailing} arm, traced here (together with the leading arm) by the dashed blue line. 
The cyan asterisk
corresponds to the position of the most bound star particle, with 
the sense of motion in the orbit shown by the arrow. 
Dark matter particles are removed from the progenitor at earlier
times, causing the stream to be dynamically old and with poor
coherence in space, although characteristic "shells" or "caustics" can
still be distinguished in the distribution.  On the other hand, star
particles are more clumped, approximately tracing the last loop ({\it
trailing} arm) and next loop ({\it leading} arm) of the orbit.  }
\label{fig:satview}
\end{figure}
\end{center}

\section{Results}
\label{sec:analysis}

It is not possible to fully constrain the orbit of the Orphan stream
with only the available radial velocity and distance
information. Proper motion measurements would be needed to that
end. However, given that the stream extends $\sim 55^\circ$ on the
sky, it is possible to find a suitable orbit by requiring that it
should pass through both ends of the observed stream, with the
measured (although preliminary) radial velocities and distances. This
can be done by requiring that the total angular momentum of both ends
of the stream be the same (for example the orientation of the orbital
plane is constrained by the cross product of the position vectors of
the stream end-points). With this condition, we randomly generate
possible orbits that match (within the errors) all observables.

The satellite is placed at one apocenter of such an orbit, from where
we follow its subsequent evolution within the Milky Way's
gravitational field. The orbit is confined to the inner 40 kpc of the
Galaxy halo (apocenter: $r_{apo}=38$ kpc, pericenter: $r_{per}=7$
kpc); where the gradient of the potential is large, producing strong
tides that fully disrupt the progenitor in less than $\sim 3.5$
Gyr. The snapshot view with the final distribution of satellite
particles after 5.3 Gyr is shown in Figure \ref{fig:satview}. The box
is 100 kpc on a side and the projection is along the x-coordinate.
The green rectangle shows the portion of the stellar trail that would
represent the Orphan stream in our model. The distribution of dark
matter (black dots) is broad and shows no spatial coherence. Due to
its more extended distribution, dark matter particles are stripped off
first, and hence this debris is dynamically older, and consequently
more diffuse. However, we can still identify by eye a few shells or
``caustics'' that outline the turn-around points along the orbit
\citep{quinn87,quinn88,helmi99}

The luminous component of our satellite is a factor $\sim17$ 
more concentrated (the ratio of their half mass radii is:
$r_h^{drk}/r_h^{str}=16.7$), which makes its core more resilient to
tidal disruption than the extended dark matter. Nevertheless, after
the second pericenter passage, the first stellar trails are produced
(at which point the satellite has lost more that 95\% of its initial
dark matter content).  The stellar streams hence grow at the expense
of the satellite's pruning, a process that, if strong and enduring enough,
drives to the irreversible disruption of the object.

Note that we have followed the evolution of the model satellite for
5.3 Gyr. This timescale is partly driven by the initial properties of
the satellite and by our desire to reproduce the current properties of
the stream. As discussed in detail in \S~\ref{ssec:mod-deg}, this
choice is not unique as strong dependencies between the initial
conditions of the model satellite and the integration time exist.
Nonetheless, the above choice of an integration timescale shorter than
a Hubble time for a relatively tight orbit may be also justified in
light of recent cosmological simulations. These have shown that
satellites are often accreted in groups \citep{li08}, and that
through multiple-body interactions acting during the tidal
dissociation of such groups their orbits can be changed significantly
\citep{sales07b,ludlow08}. Such events are relatively common, and
might explain how our model satellite was put on its current orbit
only 5.3 Gyr ago.

\begin{figure}
\centering 
\epsfxsize=8.9cm 
\epsfysize=11.5cm 
\epsffile{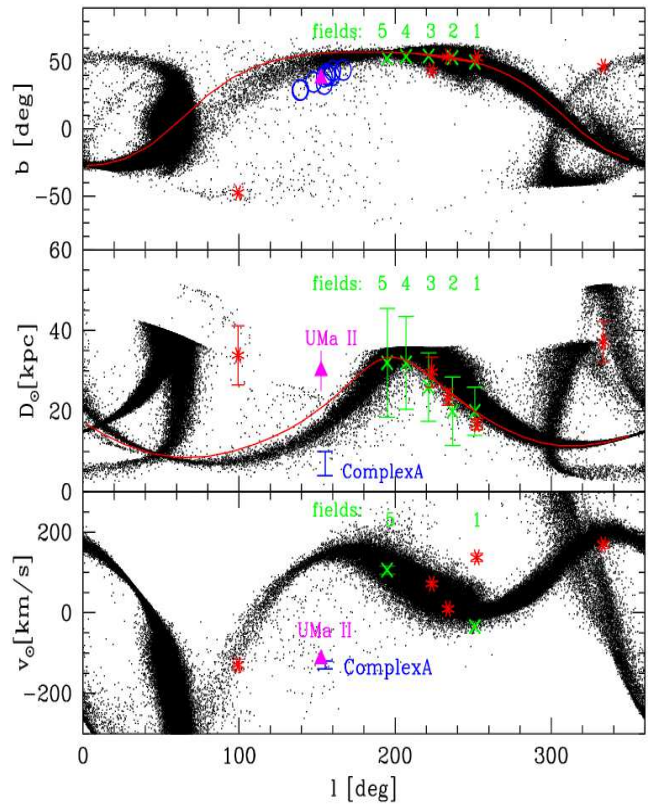}
\caption{ All sky view of the final distribution of stellar particles
in our model.  We show in the upper panel the projected positions in
galactic coordinates $(l,b)$, middle and bottom panels correspond to
the heliocentric distances ($D_\odot$) and velocities ($V_\odot$),
respectively.  The thin red line shows (a portion of) the trajectory
of a point-mass on the same orbit as the Orphan Stream progenitor,
integrated in the analytic potential of the Milky Way.  Green crosses
indicate the measurements for the Orphan Stream, fields 1-5 taken from
Belokurov et al. (2007). For comparison with previous works, we also
include current estimations for the positions, distances and
velocities for Complex A (blue empty circles) and Ursa Major II
\citep[magenta filled triangle,]{martin07b,simon07}. Red asterisks show
the properties of 5 red giant stars from the Spaghetti survey,
probably related to the stream (see \S~\ref{sec:predictions} for
further details).}
\label{fig:b_vel_d_vs_l}
\end{figure}

\subsection{The orbit and the stream global properties}
\label{ssec:orbit}

Figure \ref{fig:b_vel_d_vs_l} shows the position on the sky in
galactic coordinates, the heliocentric distances and velocities of the
star particles in our model, in comparison with the measurements of
the Orphan Stream reported by Belokurov et al. (2007b).

In this snapshot, the former center of mass of the satellite is
located at l = $306.1^\circ$, b=$13.9^\circ$, at a heliocentric distance
of 13 kpc and moving with a velocity of $v_\odot = 97.6$ km/s.

The most recent wrap of the orbit is enclosed between galactic
latitudes b=[-30$^\circ$,+60$^\circ$], and is pretty well traced by
the majority of the star particles (see upper panel in Figure
\ref{fig:b_vel_d_vs_l}). However, we can also see evidence of previous
loops, visible as a bulk of particles with l$<90^\circ$ and also at
l$>280^\circ$.  The section of the stream that matches the Orphan
Stream observations forms an arc of $\sim 55^\circ$ in the {\it
trailing} arm (the sense of motion in the orbit is pro-grade, i.e.,
towards larger galactic longitudes), where the maximum surface
brightness is reached. The leading arm is responsible for the
overdensity of particles with l$<90^\circ$ and also l$\sim
300^\circ$. The different branches are better distinguished in the
middle panel of Figure \ref{fig:b_vel_d_vs_l}, where the leading arm
has a maximum distance of $\sim 40$ kpc and the particles in the tip
of the trailing arm are located at l$>320^\circ$ with distances up to
$D_\odot\sim 50$ kpc.

Figure \ref{fig:b_vel_d_vs_l} shows that positions,
distances and velocities determined for fields 1-5 
(Belokurov et al. (2007b)) of the Orphan Stream 
are well reproduced by our model. Our distances 
might be slightly larger than measured by Belokurov
et al. (2007b) in fields 2 and 3, but still 
in good agreement, especially given the size of
the error bars. The velocities are also nicely 
consistent with the observed values, considering
that the latter were determined with barely more
than a couple of dozen stars per bin. Although some 
attempts to improve the observational estimations 
of the velocities in the Orphan Stream fields have 
already been made, unfortunately they have not yet
succeeded (Belokurov, private communication). 
Nevertheless, it is clear that a more definitive test 
to the orbit (and the progenitor) presented in this
work will come from better measurements of velocities
and smaller distance brackets for each of the fields.

A possible association between the Orphan Stream and Complex A was
suggested by \cite{belokurov07b, fellhauer07,shoko07}. Complex A is a
set of high velocity clouds whose position in the sky lies along the
best-fitting great circle traced by the Orphan Stream (see Figure 7 in
\cite{belokurov07b}). However, the model presented here shows no
connection (in the current as well as in previous loops) between these
two objects (see Fig. \ref{fig:b_vel_d_vs_l}). Our results also do not
support an association between the stream and UMa~II, as
suggested by Fellhauer et al. (2007).

\subsection{Stream intrinsic properties}
\label{ssec:moreobs}

In the following paragraphs we address how well the
intrinsic properties of the Orphan Stream
(magnitude, extension, width, etc) are 
reproduced in our model:
\begin{center}
\begin{figure}
\includegraphics[width=84mm]{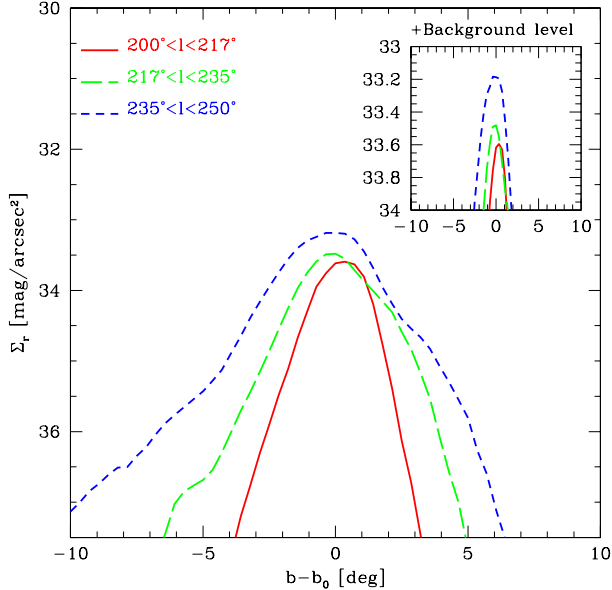}
\caption{Transverse surface brightness profile of the simulated
stream, in the region that matches
the location of the Orphan Stream. For each longitude range we
re-centred the distribution according to its baricenter latitude,
$b_0$. The main box shows the total distribution of light, while the small
indicates the likely ``detection'', i.e., the excess of flux
over a background surface brightness $\Sigma_r^F=34$ $\rm mag/\rm arcsec^2$.
As discussed in the text, accounting for the foreground and background 
contribution helps to reproduce the narrow
cross-section of the Orphan Stream shown in SDSS data.
}
\label{fig:width}
\end{figure}
\end{center}
\begin{center}
\begin{figure}
\includegraphics[width=84mm]{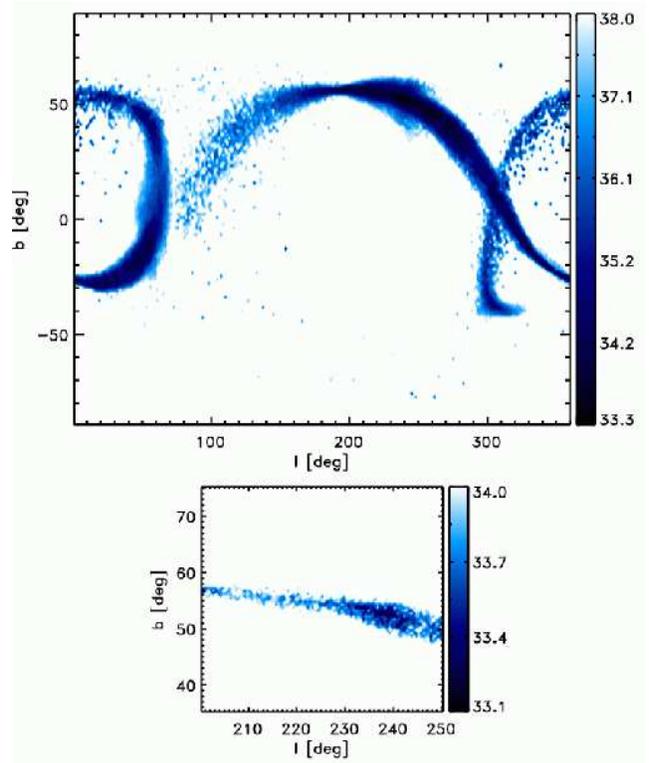}
\caption{{\it Main box:} All-sky surface brightness map of the
simulated satellite and its debris, assuming a mass-to-light ratio for
star particles of $\gamma=2.9$ (see text for details).  {\it Small
box:} zoom-in into the Orphan Stream region. Notice the change of
scale in the density map. In this box we have considered
$\Sigma_r^F=34$ $\rm mag/\rm arcsec^2$ as the lower limit in the
surface brightness, in order to account for the foreground and
background stars on the Milky Way in this region of the sky. The
average FWHM of the ``observable'' portion of the simulated stream is
then $\sim 2^\circ$ in nice agreement with the observations.  The
trend of increasing number of stars towards larger latitudes is also
reproduced in our model.  }
\label{fig:densmap}
\end{figure}
\end{center}

\begin{itemize}

\item {\it Stream width}\\ One of the peculiarities of the Orphan
Stream is its small cross section ($\sim 2^\circ$), that distinguishes
it among its siblings, e.g., the Sagittarius Stream or Monoceros Ring.
The total width of the model stream in the region l=[$200^\circ,
250^\circ$] is $15^\circ$, significantly wider than what has been
reported. However, much of this stream is at such low surface
brightness levels that it would remain undetected given the expected
background, as shown for example in Figure 5.  To account for the
contribution from background and foreground stars we use the Besancon
Model{\footnote{http://bison.obs-besancon.fr/modele/}} 4
\citep{besancon}. From this we estimate the number of stars in our
Galaxy that are expected within a 15 square degree region of the sky
centred at the Orphan Stream's position.  We apply the apparent magnitude
limit and color cuts adopted by Belokurov et al. (2007b) (stars with
$21 < r < 22$ and $(g-r) < 0.4$); obtaining an $r$-band surface
brightness for the field: $\Sigma^F_r \sim 33.8$ $\rm mag/\rm
arcsec^2$.  Therefore, the stream stars would be confused with the
background at $\Sigma_r \sim 34 \rm mag/\rm arcsec^2$.  When we apply
this threshold to our simulation, we find that the ``observable''
portion of the stream nicely reproduces the $\sim 2^\circ$
cross-section.  This is illustrated in detail in Figure 4, where we
show the average cross-section profiles of the simulated stream and
the effect of adding the foreground and background contributions. The
final structure of the stream is shown in the bottom panel of
Figure~5.

\item {\it Stream length}\\
The current detection of the Orphan stream covers an arc of 
$\sim 55^\circ$ on the sky. This strongly constrains
the time required for the debris to spread in the simulation
to a matching extent. For our model we need about $\sim$ 5 Gyr.
Shorter timespans of the modelled satellite in the fixed potential 
($t<4.5$ Gyr) generate streams that are not long enough, 
while a longer time integration ($t>6.3$ Gyr) creates  
tidal streams that are too wide and diffuse compared to the
Orphan Stream.

\item {\it Stream luminosity}\\ The total amount of mass in the
simulated stream depends on the limit in the surface brightness we
adopt. Applying the threshold given in the observations by the field
brightness, $\Sigma_r^F$, we obtain $m \sim 1.1 \times 10^5 M_\odot$.
This translates to a luminosity: $L \sim 2.3 \times 10^4 L_\odot$, or
an absolute magnitude $M_r \sim -6.4$. These numbers are in very good
agreement with Belokurov et al.'s measurements, who estimate $M_r \sim
-6.7$. Nevertheless, we notice that the real mass in the stream is
twice this number, once the limiting surface brightness considered
drops to $\Sigma_r=38 \rm mag/\rm arcsec^2$ instead of the
$\Sigma_r^F=34 \rm mag/\rm arcsec^2$ imposed by background and
foreground stars.

\item {\it Stream gradient}\\
Our model also reproduces the non-uniformity in the surface 
brightness of the Orphan Stream. As pointed out by Belokurov
et al., the density of stars noticeably increases towards
lower latitudes. Bottom panel of Figure \ref{fig:densmap} shows 
that there are approximately twice as many stars in 
$\rm l>230^\circ$ (fields 1-2) than for the upper tip 
$\rm l<220^\circ$ (fields 4-5). This effect is generated
by: i) fields 1-2 are closer to the former progenitor
than the upper tip of the stream, and ii) fields 4-5
are 50\% farther away from the Sun than the lower section of the
stream, compromising even more its detection.

\item {\it Final fate of the progenitor}\\ The large degree of
disruption in our model satellite needed to generate an arc on the sky
as large as $\sim 55^\circ$ prevents the progenitor to survive as a
self-bound entity for more than $t\sim 3$ Gyr. A robust conclusion
that we can draw after having explored several models for the
satellite, starting from conditions that resemble today's Milky Way
dwarfs, is that any remnant at the present time should be very close
to (if not already) fully disrupted.
\end{itemize}

\subsection{Model degeneracies}
\label{ssec:mod-deg}

The orbit presented here is likely to be one of many possible
ones. This is because of the freedom introduced by the lack of proper
motions measurements and the large uncertainties in the radial
velocities and distances, as well as by the unknown exact form of the
gravitational potential of the Galaxy. 

Furthermore, as a final comment in this section, we point out that the
uncertainties on the measurements and the lack of an entire sky survey
with the resolution of the SDSS, makes the problem of determining the
properties of the Orphan Stream progenitor quite degenerate (see
Fellhauer et al. (2007) for a similar conclusion). The approach in
this paper has been to choose a object that resembles one of today's
Milky Way dwarfs.  Nevertheless, the solution is not unique.

There are two factors that largely determine the success of a model on
the orbit presented here: the central density of the satellite (or the
average density within the half-light radius) and the scale of the
stellar distribution (see for instance Figure 2 of Pe\~narrubia et
al. 2008). Note that these two factors also determine the stellar
velocity dispersion, a quantity that has direct impact on the stream's
width.

The satellite's central density determines the likelihood of survival
and the degree of disruption of the object on the current
orbit. Hence, this will determine how long the streams are and for how
long the object has been orbiting the Galaxy. For example, the
observations constrain the length of the stream to an arc of 55
deg. Assuming that the Orphan Stream is actually the brightest section
of the total (full-sky) stream, and given that no progenitor has been
found close to the region where the stream reaches its peak surface
brightness, this implies that the object must be close to (if not)
fully disrupted. Therefore the initial average density of the object
must be comparable to the density of the host at the pericenter
distance ($\sim 3 \times 10^7 M_\odot$~kpc$^{-3}$). An object with a
significantly higher central density will not suffer enough tidal
stripping nor shocks to be significantly perturbed and disrupted. An
object with a much lower density will be fully disrupted giving rise
to streams that are too faint to be observable. This is why a model
like ours needs 5.3 Gyr to give rise to an object that is fully
disrupted and to streams of the required length. A model with a
slightly higher dark-matter density, but the same stellar
distribution, would survive longer, and hence the integration time
required to reach the present state would also need to be larger. A
model with a smaller amount of dark matter near the center (e.g
reduced by 50\%) is also possible, since it does not lead to
significant changes in the observables, as long as the satellite
evolves for a shorter time in the Milky Way potential ($4.7$ Gyr
instead of $5.3$ Gyr).

The second factor that is crucial in our model is the extent of the
luminous component, since this determines almost completely the
narrowness of the observed streams.  In principle, an interplay
between a shorter time integration and a more extended progenitor
could also provide a suitable model for the stream.  However, we were
unable to obtain a good match between simulations and observations by
using a satellite model as extended as the one proposed by Fellhauer
et al. (2007).  These authors used a two-component model with similar
mass for the luminous particles, but its half light radius is $r_h \sim
480$ pc, approximately 3.5 times larger than our fiducial
progenitor. Such an extended object in our preferred orbit produces a
stream that is too wide and with too low surface brightness in
comparison to the Orphan Stream. On the other hand, the mass in the
luminous component could also be increased maintaining the mass-size
relationship observed for the local dwarf spheroidals. However, these
more extended progenitors give rise to tails that are too wide and
(now) too bright to fit the Orphan Stream observations.

As stated above, these degeneracies are important since the
integration timescale can then be made as long as desired by
increasing the average density of the progenitor. For example, one may
desire a model which has been on the present orbit around the Galaxy
for the past 10 Gyr. By performing numerical experiments we have found
that a satellite with a mean density within the half light radius
$r_h$ which is 10 times larger than our fiducial model (i.e., $\sim
2.9 \times 10^9 M_\odot$~kpc$^{-3}$) will achieve after 10 Gyr the
level of disruption needed to reproduce the observations. Such a model
can be obtained by increasing the dark matter halo mass by a factor 10
while keeping its scale-length $a_{sat}$ fixed. However, this will
lead to a significantly larger stellar velocity dispersion in
comparison to the previous model (with $\sim 7$ km/s), and hence will
fail in reproducing the width of the Stream. Therefore also the extent
of the luminous component must be modified to become more concentrated
than our fiducial model. The magnitude of this modification can be
derived from the following argument based on the initial stellar
velocity dispersion of the system $\sigma_{str}$. Since most of the
mass is in dark matter, we can neglect the contribution of the stars
to the velocity dispersion, and therefore assuming, as before, a
Hernquist halo and a Plummer profile for the stars we find:
$$ \sigma_{str}(r)= G M_{drk} (r^2+b_{sat}^2)^{5/2} \int^\infty_{r}
\frac{1}{(x^2+b_{sat}^2)^{5/2}} \frac{dx}{(x+a_{sat})^2}$$ where
$a_{sat}$ and $b_{sat}$ are the characteristic radii of the Hernquist
and the Plummer profiles, respectively \citep{hernquist93}. The
dark halo mass, $M_{drk}$ is fixed by our requirement of a given
central density. If we now vary $b_{sat}$ we can get a range of
velocity dispersions that goes from 4 km~s$^{-1}$ if $b_{sat}=20$~pc
up to $\sim 24$ km~s$^{-1}$ for $b_{sat}=0.1$~kpc as before.  If we
use our fiducial model as a guide, the new model integrated for 10 Gyr
will be successful if $\sigma_{str} \sim 7$~km~s$^{-1}$ after $\sim 5$
Gyr of evolution. Since disruption causes the velocity dispersion to
drop, models with initial $\sigma_{str}$ in the range
9--12~km~s$^{-1}$ will generate the kind of progenitor we are looking
for. According to the equation above, that corresponds to a half light
radius $r_h \sim$ 50-65 pc.

This new progenitor should also fulfil the constraints in luminosity
given by the brightness of the visible portion of the Orphan Stream
(see Section 2), which implies $L \ge 10^5 L_\odot$. Therefore, we
find that by fixing the central density, and changing the parameters
of the luminous component accordingly, we can generate a family of
progenitors that are roughly consistent with today dwarfs satellites.
On this basis, we conclude that models evolving for 5 to 10 Gyr and
with the properties specified above could represent feasible
candidates to the progenitor of the Orphan Stream.

For completeness, we have also explored single-component (no dark
matter) models (reminiscent of a globular cluster or a dwarf galaxy
that has already lost its subhalo) as tentative progenitors for the
stream. We find that such objects should be more concentrated
(typically half light radius $\sim 40$ pc) and need to be integrated
for a longer time ($\sim 6.5$ Gyr) than our fiducial model, in order
to reproduce the length and width of the Orphan Stream. As discussed
above, a higher density progenitor could be integrated for a longer
timescale leading to a similar configuration. It is interesting to
note that an object like a globular cluster has too high a central
density ($\sim 10^{12} M_\odot$~kpc$^{-3}$) to be fully disrupted within
a Hubble time on this orbit. The amount of mass in the currently
detected stream is fixed from observations, hence, the total
luminosity of these models should also be $L > 2 \times 10^5 L_\odot$,
as in the double-component case. Progenitors with such properties
could be considered unlikely given the properties of the satellites in
the Local Group: they are too bright and extended to be a globular
cluster and too compact to represent a Milky Way dwarf galaxy (see for
instance Fig. 10 of Pe\~narrubia et al. 2008).  Therefore, although
we cannot rule out a single-component progenitor with such
characteristics (since today's satellites might not be representatives
of earlier accreted ones); the lack of observations of similar objects 
turns the dark-matter dominated model presented before a more suitable
candidate for our analysis.

\section{Preliminary predictions and future detection prospects}
\label{sec:predictions}

The upper panel of Figure $\ref{fig:densmap}$ shows a dense enough
section of the stream, likely identifiable with the current sensitivity
of the SDSS; although unfortunately, sitting outside
its coverage area. This new stretch contains about $\sim$2 times
less stars and traces an arc of $\sim 35^\circ$ following the current
Orphan Stream detection. It extends from $\rm l=250^\circ$ to $\rm
l\sim 290^\circ$ in the North galactic hemisphere. The leading arm
could in principle be also identified in a survey covering the
southern galactic hemisphere.  However, its position towards the
galactic center ($\rm l=[-60^\circ,70^\circ]$) and its lower latitude
($\rm b=[-30^\circ,10^\circ]$) might compromise the chances of
positive imaging detection. In addition to this, its mean surface
brightness is on average lower than that of the trailing arm, where
the Orphan Stream sits.  This difference can be attributed to the fact
that the portion of the trailing arm identified as the Orphan Stream
lies close to apocenter of the orbit, where the spacial coherence of
particles is locally enhanced (e.g., formation of shells or caustics
seen in Figure \ref{fig:satview}) reaching the maximum surface
brightness across the full sky.

An interesting alternative to these photometric approaches, is offered
by studies involving also stellar kinematics.  Stars remain coherent
in velocity space as a consequence of the conservation of phase-space
density \citep{helmi99}, enhancing the correlations of their dynamical
parameters.

One relevant study of stellar kinematics of (outer) halo stars is
provided by the Spaghetti Survey. This pencil-beam survey of
high-latitude fields, first described in Morrison et al. (2000),
\nocite{morrison00} uses solely red giants as distant halo
tracers. The final dataset consists of 102 spectroscopically confirmed
giants (Starkenburg et al.\ 2008, in preparation). We find five giants
within this dataset that could be related to the stream in our model
(shown by the red asterisks in Figure \ref{fig:b_vel_d_vs_l}). Three
of them are located right on top of the Orphan Stream fields as
detected in Sloan. For two of these, the projected positions,
distances and also heliocentric velocities in good agreement with the
simulations, indicating a large probability of a real physical link to
the stream. The case for the third giant is less clear, since its
position agrees with the model, but its velocity is too large (given
the radial velocities measured by \cite{belokurov07b} are still
preliminary, it is possible that this giant is physically related to
the Orphan Stream, but not in our model). The remaining two giant
stars could belong to different wraps, although they sit on relatively
low-density portion of the stream. Table \ref{tab:giants} quotes the
positions, velocities and metallicities of the selected giants
measured by the Spaghetti survey, together with our assessment of the
likelihood of membership to the Orphan Stream progenitor according to
our simulations.

While red giant surveys like the Spaghetti Survey may be suitable to
find stream members and constrain the stream's properties such as
radial velocity, the number of detectable candidates is strongly
limited by the surface brightness of the stream. The approximate
surface brightness of $\sim32.4$ mag arcsec$^{-2}$ (Belokurov et
al. 2007b) transforms to barely 1.3 giants per square degree
\citep{morrison93}. With more fields on the stream and a more careful
analysis of the existing SDSS spectra for giant candidates in fields
on the stream, it should be possible to constrain the radial
velocities better. However, even fainter substructures might need
other tracers such as main sequence stars to determine their
properties.

Note also that some sections of the stream in our simulations fall
relatively close to the sun (heliocentric distances of the order of 5
kpc) and could, in principle, be accessible to RAVE
\citep{steinmetz06}. Measurements of the proper motions of the stars
associated to the stream could also be used to falsify our model.  The
orbit presented here as $(\mu_\alpha,\mu_\delta)\sim (-0.10,-2.42) \rm
mas/yr$ at $\alpha = 162^\circ$ and $\delta = 0^\circ$, i.e. on Field 1. These are
within reach of future astrometric missions like GAIA.
   
\begin{table*}
\caption{ Galactic coordinates, distances, velocities and
metallicities of the Spaghetti Survey giants shown in Figure
\ref{fig:b_vel_d_vs_l}. Errors in the quantities are also quoted. The
last column denotes how consistent the membership to the Orphan Stream
progenitor is according to our model.  }
\begin{tabular}{|c|c||c|c|c|c|c|c|c|c|}
\hline
Location & $l$ & $b$ & $D_\odot$ & $\Delta_{ D_\odot}$  
& $V_\odot$ & $\Delta_{V_\odot}$ & $Z$ & $\Delta_{Z}$ & Membership \\
& (deg)& (deg) & (kpc) & (kpc) & (km/s) & (km/s) & (dex) & (dex) &  \\
\hline
Field 3 & 223.45 & 43.44 & 29.71 & 3.71 & 71.8 & 5.5 & -1.38 &0.27 & likely \\
Field 2 & 233.83 & 53.71 & 23.46 & 2.03 & 9.6 & 7.5 & -1.24 & 0.26 & likely \\
Field 1 & 252.28 & 52.98 & 16.72 & 1.39 & 137.1 & 10.1 & -1.24 & 0.26
& unlikely\\
Older wrap & 99.38 & -47.59 & 33.88 & 7.37 & -129.9 & 14.& -1.60 & 0.59 & possible\\
Older wrap & 333.34 & 46.50 & 37.16 & 5.10 & 168.7 & 10.1 & -1.72 &
0.27 & possible\\ \hline
\end{tabular}
\label{tab:giants}
\end{table*}
%

\section{SUMMARY AND CONCLUSIONS}
\label{sec:conc}

We have presented N-body simulations that reproduce the properties of
the recently discovered Orphan Stream.  We have studied an orbit able
to match, the positions, velocities and distances of the stream.  The
orbit is in good agreement with the continuous nature of the stream,
which may well be a single wrap of the trailing arm of a fully
disrupted satellite.  The main features of the stream, such as length,
cross-section, luminosity and surface brightness are also recreated
successfully in the model.  However, the orbit does not provide an
obvious association of the stream with Complex A or Ursa Major II, as
had been suggested previously by \citet{shoko07} or Fellhauer et al. (2007).

The satellite in our model consists of two components: an extended
dark matter halo within which the stars are deeply embedded. We find
that the Orphan Stream progenitor could have been an object that
resembles one of today's Milky Way dwarfs, such as Carina, Leo~II,
Draco or Sculptor, and hence, likely to be dark matter dominated.  The
model predicts the total disruption of this object in order to
generate the $\sim 55^\circ$ stellar stream needed to match the
observations. This indirectly implies that objects with the properties
of the above-mentioned dwarfs can only survive as self-gravitating
systems today, because they populate the outer regions of the Galaxy
potential, where the gravitational forces are smaller, and the
dynamical times of the orbits are longer.

The Orphan Stream distinguishes itself from the previously known tidal 
features (e.g., Sagittarius Stream or Monoceros Ring in our Galaxy,
or the Giant Arc in Andromeda) by its small cross-section. 
With an average width of $\sim 2^\circ$ it is about $\sim5$ times
narrower than the Sagittarius Stream. However, as argued above, 
this does not imply that its progenitor is an exceptional object.
Such streams may therefore be far more common than suspected thus far. 
The visible strip of the stream may also be just the ``tip of the iceberg'', 
where the amount of mass under our detection limit could
be at least twice the current estimation.

We conclude that additional discoveries of tidal features such as the
Orphan Stream are fundamental to our understanding of the outer halo of 
galaxies, as well as to tests of current cosmological
models. The detectability of such coherent but very faint structures
is compromised not only by the initial luminosity of their
progenitors, but also by the dynamical ages of these structures, since
they become fainter as they age and diffuse away.  The design of
techniques to recover these ghostly features needs to be carefully
thought out, in order to enable the detection of streams from objects
originally populating the faint end of the luminosity function. Future
panoramic surveys of the Milky Way stars, as well as the determination
of their kinematics, ages and metallicities will play a fundamental
role in the Galactic astronomy of the next decades, helping to unveil
the secrets recorded in the stellar halo of our own Galaxy.

\section*{Acknowledgements}
\label{acknowledgements}

LVS would like to thank Donald Lynden-Bell, Vasily Belokurov, Nicolas Martin, 
Mario G. Abadi and Julio F. Navarro for very helpful discussions,
as well as to Vibor Jeli\'c for his help with IDL macros.
LVS and AH gratefully acknowledge NWO and NOVA for financial support.
HLM was supported by NSF grants AST-0098435 and
AST-0607518.
EO is partially supported by NSF grants AST-0205790 and AST-0507511.
T.S acknowledges funding from  Physics Frontiers Center/Joint 
Institute for Nuclear Astrophysics
(JINA),  awarded by the U.S. National Science Foundation.
We also thank the anonymous referee for useful suggestions and
comments that helped to improve the previous version of the paper.


\end{document}